\documentclass{article}

\usepackage{arxiv}

\usepackage[utf8]{inputenc} 
\usepackage[T1]{fontenc}    
\usepackage{hyperref}       
\usepackage{url}            
\usepackage{booktabs}       
\usepackage{amsfonts}       
\usepackage{nicefrac}       
\usepackage{microtype}      
\usepackage{lipsum}
\usepackage{graphicx}

\usepackage{amsmath}
\usepackage{esint}
\usepackage{subcaption}
\usepackage{comment}

\usepackage{booktabs}
\usepackage{amssymb}
\usepackage{mhchem}
\usepackage[colorinlistoftodos]{todonotes}
\usepackage{dcolumn}
\usepackage{tabularx}
\usepackage[export]{adjustbox}
\usepackage{makecell}
\usepackage{amsfonts}       
\usepackage{nicefrac}       
\usepackage{microtype}      
\usepackage{lipsum}
\usepackage{tipx}
\newcolumntype{C}[1]{>{\centering\let\newline\\\arraybackslash\hspace{0pt}}m{#1}}

\usepackage{nameref}
\usepackage{cleveref}
\usepackage{nomencl}
\makenomenclature

\title{Lessons in Reproducibility: Insights from NLP Studies in Materials Science}

\usepackage{authblk}
\author[1*]{Xiangyun Lei}
\author[2,5]{Edward Kim}
\author[1,3]{Viktoriia Baibakova}
\author[1,4*]{Shijing Sun}

\affil[1]{\footnotesize Toyota Research Institute, Los Altos, CA 94022}
\affil[2]{\footnotesize Cohere, Toronto, Canada}
\affil[3]{\footnotesize Lawrence Berkeley National Laboratory, Berkeley, CA 94720}
\affil[4]{\footnotesize University of Washington, Seattle, WA 98195}
\affil[5]{\footnotesize University of Toronto, Toronto, Canada}

\begin{document}
\maketitle
\begin{abstract}
Natural Language Processing (NLP), a cornerstone field within artificial intelligence, has been increasingly utilized in the field of materials science literature. Our study conducts a reproducibility analysis of two pioneering works within this domain: "Machine-learned and codified synthesis parameters of oxide materials" by Kim et al., and "Unsupervised word embeddings capture latent knowledge from materials science literature" by Tshitoyan et al. We aim to comprehend these studies from a reproducibility perspective, acknowledging their significant influence on the field of materials informatics, rather than critiquing them. Our study indicates that both papers offered thorough workflows, tidy and well-documented codebases, and clear guidance for model evaluation. This makes it easier to replicate their results successfully and partially reproduce their findings. In doing so, they set commendable standards for future materials science publications to aspire to. However, our analysis also highlights areas for improvement such as to provide access to training data where copyright restrictions permit, more transparency on model architecture and the training process, and specifications of software dependency versions. We also cross-compare the word embedding models between papers, and find that some key differences in reproducibility and cross-compatibility are attributable to design choices outside the bounds of the models themselves. In summary, our study appreciates the benchmark set by these seminal papers while advocating for further enhancements in research reproducibility practices in the field of NLP for materials science. This balance of understanding and continuous improvement will ultimately propel the intersecting domains of NLP and materials science literature into a future of exciting discoveries.
\end{abstract}

\section{Introduction}
Natural Language Processing (NLP), a dynamic subfield of artificial intelligence, has revolutionized how computers understand, interpret, and generate human language. This powerful technology, encompassing models ranging from the simplistic Bag of Words to the more advanced architectures such as Bidirectional Encoder Representations from Transformers (BERT) \cite{BERT} and Generative Pre-trained Transformer (GPT) \cite{openai2023gpt4, brown2020language}, has found successful deployments across a broad range of applications, from search engines and voice assistants to machine translation and knowledge extraction.

Recently, these NLP models have been harnessed to streamline research in the realm of materials science literature \cite{olivetti2020data}. Researchers have custom-trained these models to delve into the vast corpus of materials science literature, thereby facilitating the extraction of hidden patterns and accelerating scientific innovation. Through the automation of data extraction and analysis, these NLP models catalyze interdisciplinary collaboration, promote the discovery of novel materials \cite{lee2023natural}, and provide valuable insights for efficient resource allocation. The intersection of NLP and materials science holds the potential not only to expedite the discovery process but also to usher novel materials and technologies swiftly to market \cite{pyzer2022accelerating}. Despite the burgeoning growth of this field, underscored by the proliferation of research articles, a critical aspect often overlooked is the reproducibility of these studies. To address this concern, our work focuses on the reproducibility analysis of two seminal publications in the field of NLP applied to materials science literature.

The first publication, "Machine-learned and codified synthesis parameters of oxide materials" by Kim \textit{et al}. \cite{Kim2017}, published in \textit{Scientific Data} in 2017, employed NLP tools to extract synthesis parameters from over 76,000 relevant research articles. The methodologies and datasets introduced in this work have significantly contributed to the community, as evidenced by approximately 100 citations in various studies. The second publication is "Unsupervised word embeddings capture latent knowledge from materials science literature" by Tshitoyan \textit{et al}. \cite{mat2vec}, published in \textit{Nature} in 2019. This groundbreaking work applies NLP to distill materials science-related knowledge from the abstracts of more than three million research articles, resulting in a pioneering model, Mat2Vec. The impact of this research is unmistakable, having received over 400 citations since its publication.

In our analysis, we reviewed the information and codebases provided in both papers and endeavored to replicate and reproduce the reported results to the greatest extent possible. Replication in this context refers to using the provided machine learning models, testing them against the same test cases referenced in the papers, and ensuring the results align. Reproduction, however, necessitates retraining and possibly rebuilding the reported models from scratch. In this study, we document each step of our procedure, offering  feedback and recommendations along the way, thereby contributing to the ongoing conversation on research reproducibility in the field. We also evaluate a cross-comparison of the NLP models in the two studies, as they both use Word2Vec models via the same underlying software library \cite{gensim}. The cross-comparison efforts serve as a benchmark for assessing the generalizability of the reported models to applications beyond the original dataset used for their model development.

\section{Paper 1: Kim et al., 2017}

In their study published in 2017, Kim \textit{et al}. addressed a significant gap in materials science by devising a comprehensive, autonomously compiled database for oxide material synthesis planning, and it has since been used for developing data-informed synthesis strategies. The authors achieved this by harnessing the capabilities of NLP, particularly the Word2Vec model\cite{word2vec}, to extract synthesis information from over 76,000 previously published research articles with oxide material synthesis information. 

Word2Vec is an NLP model that can transform words into numerical vectors when trained, capturing semantic relationships between words. Consequently, it facilitates the understanding of context and the detection of synonyms and antonyms, thereby enabling more advanced operations like "king" - "man" + "woman" = "queen". This capability has proven instrumental in the task of autonomous information extraction. For greater precision in material synthesis information extraction, the authors adopted a two-stage training process. Initially, the Word2Vec model was pre-trained on 640,000 unlabeled full-text articles on materials synthesis to learn accurate vector representations of domain-specific terms. Then another supervised model was trained using the Word2Vec model as a featurization step, and this combination of models was used to categorize words in articles into "material", "condition", "operation", etc. using 20 manually annotated articles. 

The synthesis information was subsequently extracted using a heuristic-rule guided analysis of the higher-level relationship between words. This analysis employed a grammatical parser's outputs, with relationships between chunks determined from parse tree dependencies, and word-order proximity serving as a secondary measure. The extracted synthesis parameters were then filtered and compiled into the dataset, offering a comprehensive robust resource for the synthesis of oxide materials.

The authors included two Github repositories. The first one, referred in this study as the Model repository, hosts the pre-trained Word2Vec model implemented with the gensim python package \cite{gensim}. The second repository, which we refer to it as the Plot repository, contains a tutorial Jupyter notebook explaining how the figures in the paper were generated. Here, we run the scripts provided in both repositories in the attempt to reproduce the work presented in this study. The goals are to better understand how the results were obtained and to test the model developed in this study on other materials systems beyond the examples given in the manuscript.

In both the Model and Plot repositories, the authors provided explanations and instructions in the README.md files. The installation procedure provided runs successfully with no bugs (tested on both Ubuntu and MacOS machines), and we are able to load the pre-trained Word2Vec model shared by the authors. A sample script for testing the pre-trained model is also provided in the model repository. With the script, we are able to replicate the results presented by the authors. Notably, the database identified $Li_{4}Ti_{5}O_{12}$ and six other lithium-containing oxides as the most similar materials to $LiFePO_{4}$ (Table \ref{tab:paper1_simlarity}). It is worth mentioning that 'LFP', a commonly used abbreviation for lithium iron phosphate, was also recognized as one of the most similar materials to $LiFePO_{4}$, with a similiarty score of 0.667. This finding suggests that the model is capable of identifying the similarity between $LiFePO_{4}$ and LFP, as one would expect from a Word2Vec model, which captures context-based similarity and thus tends to work especially well for capturing synonyms. However, the model were not trained to tell that LFP and $LiFePO_{4}$ refer to the same compound. We also tested the code with a new word $CsPbI_{3}$, which was not included in the example script, and the output was scientifically meaningful (Table \ref{tab:paper1_simlarity}). Additionally, we were able to reproduce other functionalities of the model, such as detecting outlier processing conditions and determining the similarity metric between two known materials by providing their names in letters instead of chemical formulas (Table \ref{tab:paper1_simlarity2}). Interestingly, we observed that whether the first letter is capitalized makes a difference in similarity prediction, an effect that was not included in the original study. This is likely due to choices in the text preprocessing for the model, as it is reasonable to preserve the upper and lower casing for the text that contains chemical formulas, where the casing carries semantic meaning (e.g., $LiCo_{3}$ is not the same as $LiCO_{3}$).

\begin{table}[!ht]
    \centering
    \caption{Reproduced results using the Word2Vec model for identifying the most similar materials given reference material. The first column is the provided sample output using the reference material "$LiFePO_{4}$". The second column is the reproduced results using the same material. The third column is the result with $CsPbI_{3}$ as the reference material, which was not included in the original work}
    \begin{tabular}{|c|cc|cc|cc|}
    \hline

        \textbf{Catogories} & \multicolumn{2}{c|}{Example provided:LiFeO4} & \multicolumn{2}{c|}{Example replicated: LiFeO4} & \multicolumn{2}{c|}{New example: CsPbI3}  \\ \hline
        \textbf{No.} & Materials & Similarity & Materials & Similarity &  Materials & Similarity \\  
        \textbf{1} & Li4Ti5O12 & 0.768 & Li4Ti5O12 & 0.768 & CaVO3 & 0.703 \\ 
        \textbf{2} & LiMn2O4 & 0.756 & LiMn2O4 & 0.756 &  CH3NH3PbBr3 & 0.703 \\ 
        \textbf{3} & LTO & 0.714 & LTO & 0.714 &  SrVO3 & 0.670 \\ 
        \textbf{4} & LiCoO2 & 0.707 & LiCoO2 & 0.707 &  BiCuOSe & 0.656 \\ 
        \textbf{5} & LiMnPO4 & 0.696 & LiMnPO4 & 0.696 & CsPbBr3 & 0.655 \\ 
        \textbf{6} & FePO4 & 0.682 & FePO4 & 0.682 &  wurzite & 0.654 \\ 
        \textbf{7} & LFP & 0.667 & LFP & 0.667 &  CH3NH3PbI3 & 0.649 \\ 
        \textbf{8} & LiNi0.5Mn1.5O4 & 0.662 & LiNi0.5Mn1.5O4 & 0.662  & CZGS & 0.647 \\ 
        \textbf{9} & FeF3 & 0.658 & FeF3 & 0.658 &  Cu2ZnSnSe4 & 0.644 \\ 
        \textbf{10} & LiV3O8 & 0.658 & LiV3O8 & 0.658 &  In4Se3 & 0.642 \\ \hline
    \end{tabular}
    \label{tab:paper1_simlarity}
\end{table}

\begin{table}[!ht]
    \centering
    \caption{Reproduced results using the Word2Vec model for calculating the similarity between two materials. the reference similarity is provided by the original work}
    \begin{tabular}{|c|cccc|}
    \hline

         & Material 1 & Material 2 & Similarity & Reference similarity  \\ \hline
        \textbf{1} & titania & zirconia & 0.599 & 0.599  \\ 
        \textbf{2} & anatase & rutile &  0.847 &   \\ 
        \textbf{3} & magnesia & lime & 0.577 &   \\ 
        \textbf{4} & hermatite & corundum &  0.524 &   \\ \hline
        \textbf{5} & Titania & Zirconia & 0.668 &   \\ 
        \textbf{6} & Anatase & Rutile &  0.798 &   \\ 
        \textbf{7} & Magnesia & Lime & 0.635 &   \\ 
        \textbf{8} & Hermatite & Corundum &  0.626 &   \\ \hline

        \hline
    \end{tabular}
    \label{tab:paper1_simlarity2}
\end{table}

Although the scripts and pre-trained model snapshot offer adequate information to evaluate some example applications of the model, the code or details for the Word2Vec model's architecture and training process are not provided. Moreover, the training data used to train the model is not openly available, likely (and understandably) due to publisher policies from which the data was obtained. Therefore, we were not able to reproduce the Word2Vec model from scratch. Further, implementations of the rest of the workflow are not provided in the repository, so we don't really know how they are coded and applied, hence not able to replicate or reproduce the complete workflow.

In the Plot repository, the authors included a notebook that presents a guide on how to generate the figures presented in their study. Although the code was originally written in Python 2, we were able to execute it in Python 3 with minor changes. This underscores the importance of explicitly indicating the programming language used in the repository. The authors made the data available in JSON format through an online data repository (figshare), and we were able to replicate all the figures presented in the notebook. However, it appears that the data required to generate the plots were hardcoded within the script, meaning that we could only reproduce the process of generating the figures but not the results themselves. Therefore, we cannot verify the accuracy of the results presented in the paper.

Overall, the substantial contribution of this work to the field is unquestionable. The primary focus is to construct the database of synthesis parameters for oxide materials, and the authors have made a commendable effort to ensure that their workflow can be replicated and built upon by other researchers. This is evident from their provision of the workflow structure, the pre-trained Word2Vec model, scripts for figure generation, and detailed Readme files in the repositories. Using the information at hand, we were able to replicate a portion of the results, which speaks to the value of the authors' approach and their commitment to reproducibility. From the perspective of enhancing reproducibility in future work, there are areas that could be further clarified to enhance reproducibility. For one, the paper discusses multiple models, but more explicit descriptions of these models and their hyperparameters could be beneficial. For instance, the authors employed a binary logistic regression classifier from the Scikit-learn library, and while the model is described, the inclusion of details such as decision boundary and other hyperparameters could aid understanding. Similarly, the use of a pre-trained Word2Vec model followed by training of a baseline and a human-trained neural network is mentioned, but the absence of detailed descriptions of the network layers and the training hyperparameters leaves room for further clarification. The inclusion of such details could promote a deeper understanding of the methods employed and facilitate smoother transitions for those building upon this work. On the other hand, it might be beneficial if the scripts provided for reproducing the plots in the paper could start directly from the database, rather than from hard-coded values.

\section{Paper 2: Tshitoyan el al., 2019}

Although the second paper by Tshitoyan el al. \cite{mat2vec} uses the same tool, namely the Word2Vec model, the goal of the work was different. In this work, the authors harnessed the power of the Word2Vec model to encode comprehensive knowledge about various materials reported in the scientific literature and showed that the embeddings could be used to make useful predictions. To achieve this, the model was trained on the abstracts of over three million materials science-related research articles. Due to the nature of the model, carefully designed application of domain knowledge is needed to apply the model and extract meaningful information. 

In this work, the authors conducted three key studies: (1) they plotted and performed simple arithmetic operations on the Word2Vec embeddings, thus demonstrating that physical knowledge was indeed encoded within the model; (2) they leveraged similarity scores to identify potential thermoelectric materials, demonstrating a novel method to get valuable insights from the model; and (3) they trained and tested models on articles published until various points in time, to underscore that the model, informed by existing literature, could reliably guide future research. Hence, the authors demonstrated that the Mat2Vec model can efficiently encapsulate materials science knowledge from the literature without the need for human supervision or labeling. The unsupervised model can anticipate functional materials years before their actual discovery, suggesting that a wealth of latent knowledge about future discoveries is nestled within past publications. The paper offers evidence for these claims by detailing multiple use cases of the mat2vec model through illustrative figures and text. 

\begin{figure}[ht]
    \centering
    \begin{subfigure}{0.45\textwidth}
        \centering
        \includegraphics[width=\textwidth]{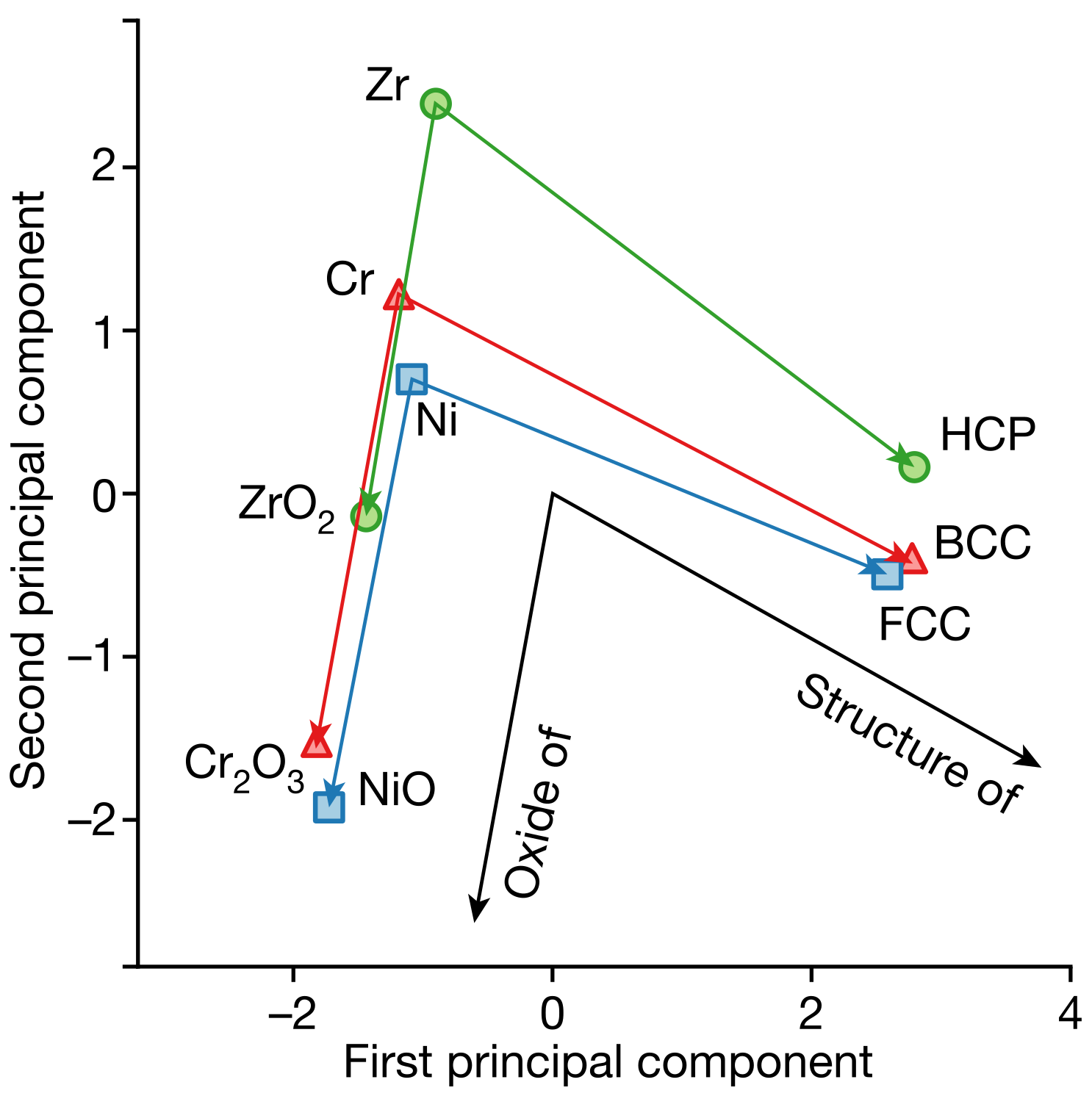}
        \caption{Original plot}
        \label{fig:paper2_arithmetic_sub1}
    \end{subfigure}
    \begin{subfigure}{0.45\textwidth}
        \centering
        \includegraphics[width=\textwidth]{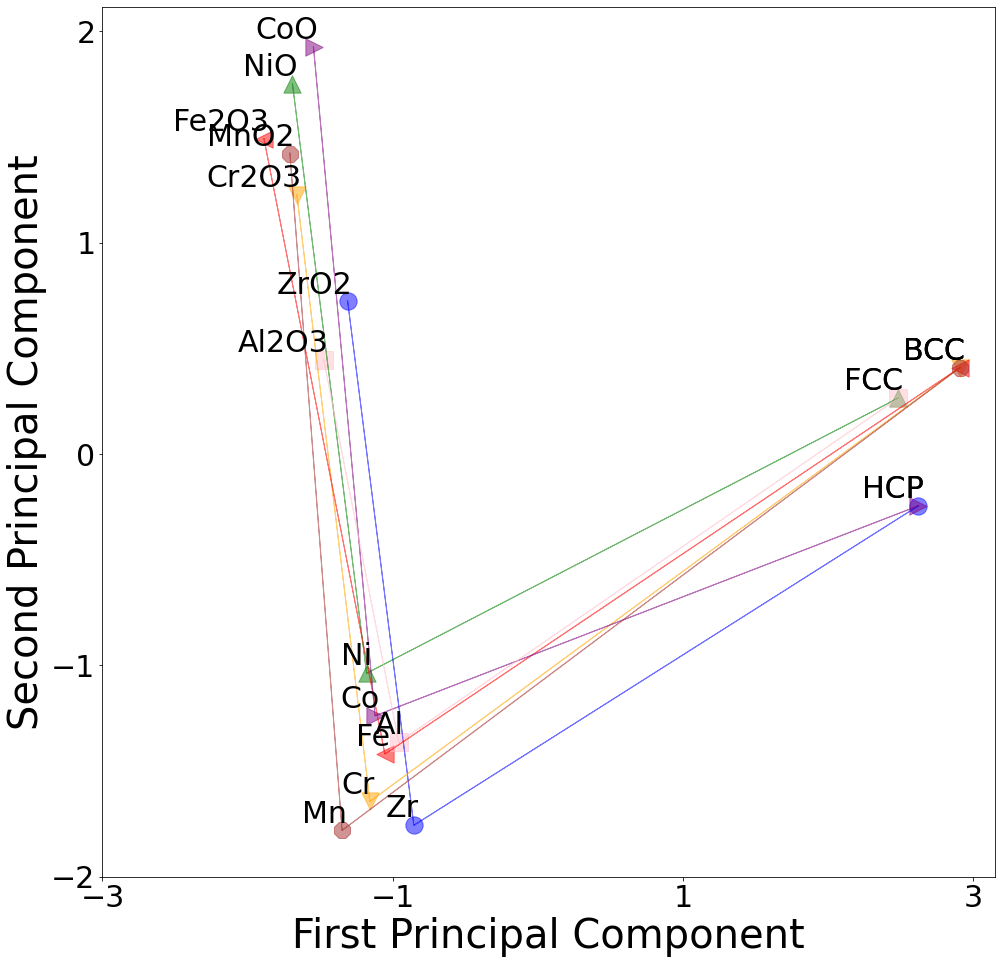}
        \caption{Repoduced plot }
        \label{fig:paper2_arithmetic_sub2}
    \end{subfigure}
    \caption{Reproducibility study of word relationships based on the predicted word embedding using the Mat2Vec model. The original plot (a) is taken from the original paper \cite{mat2vec} (Figure 1b). Word embeddings for Zr, Cr, and Ni, their principal oxides, and crystal symmetries (at standard conditions) projected onto two dimensions using principal component analysis and represented as points in space. The relative positioning of the words encodes materials science relationships, such that consistent vector operations exist between words that represent concepts such as ‘oxide of ’ and ‘structure of’. Plot (b) is our reproduction but with more examples.}
    \label{fig:paper2_arithmetic}
\end{figure}

\begin{figure}[ht]
    \centering
    \begin{subfigure}{0.9\textwidth} 
        \centering
        \includegraphics[width=0.5\textwidth]{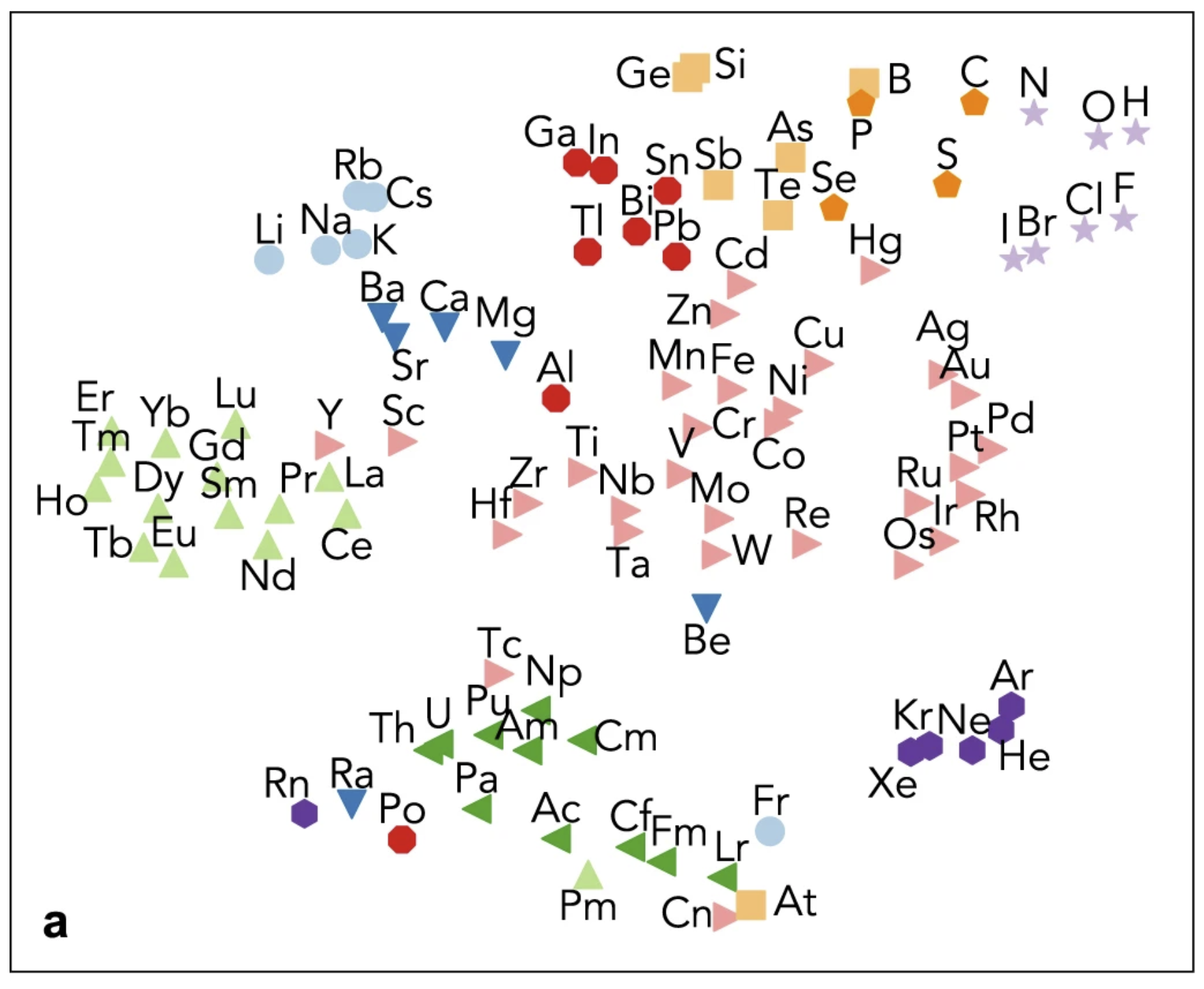}
        \caption{Taken from the original paper \cite{mat2vec}. Two-dimensional t-distributed stochastic neighbor embedding (t-SNE) projection of the word embeddings of 100 chemical element names (for example, ‘hydrogen’) labeled with the corresponding element symbols and grouped according to their classification. Chemically similar elements are seen to cluster together, and the overall distribution exhibits a topology reminiscent of the periodic table itself}
        \label{fig:paper2_element_embedding_sub1}
    \end{subfigure}
    \vskip\baselineskip
    \begin{subfigure}{0.45\textwidth}
        \centering
        \includegraphics[width=\textwidth]{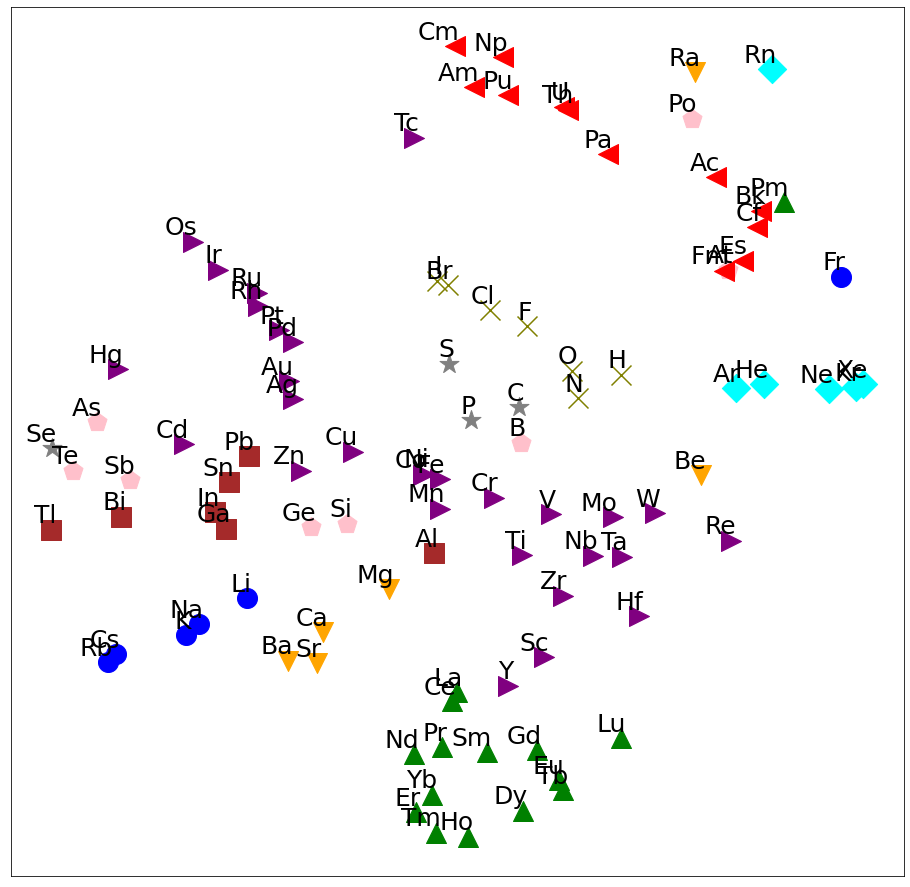}
        \caption{Reproduction of (a), also based on the predicted chemical element name embedding}
        \label{fig:paper2_element_embedding_sub2}
    \end{subfigure}
    \begin{subfigure}{0.45\textwidth}
        \centering
        \includegraphics[width=\textwidth]{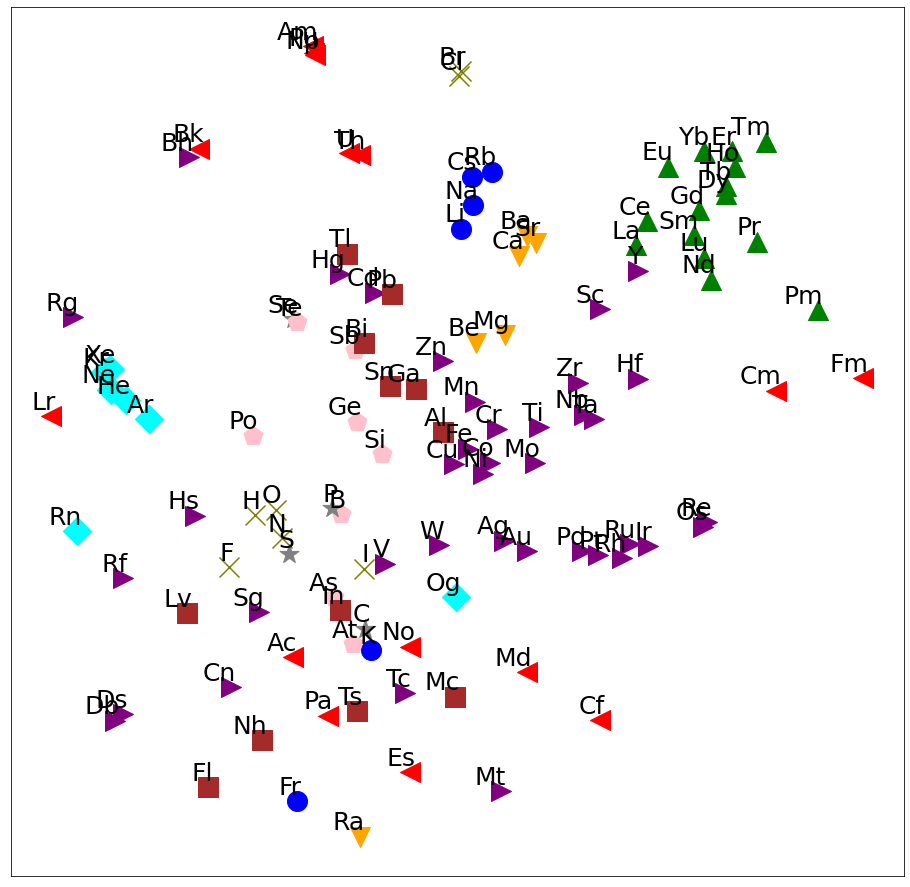}
        \caption{Same t-SNE plot as (a), but based on the embeddings of chemical symbols (for example, ‘H’) }
        \label{fig:paper2_element_embedding_sub3}
    \end{subfigure}
    \caption{Reproducibility study of elemental embedding using the Mat2Vec model. Two reproducibility studies are conducted: one based on the chemical element names (b), and one based on the chemical symbols (c).}
    \label{fig:paper2_element_embedding}
\end{figure}

Here, we try to replicate and reproduce the studies mentioned in the paper, and also extend the studies to other materials. We started by trying to install the Mat2Vec package, which is provided as supplemental material for the paper and open-sourced on GitHub along with pre-trained models (the Word2Vec model trained to abstracts of research articles). The codebase is well-documented and in our opinion, straightforward to follow. A tutorial for installing and testing the package is also provided in the Readme file, and we found no issues installing the package following the steps (on both Ubuntu and MacOS machines). Further, the pre-trained models can be loaded with no error, and the test results of the similarity study given in the paper and tutorial can be reproduced with only minor numerical noise (Table \ref{tab:paper2_simlarity}). The figure for demonstrating the word relationship based on the embedding (Figure 1b of the original paper) (Figure \ref{fig:paper2_arithmetic}), and the figure for elemental embedding (Figure S1a of the original paper) (Figure \ref{fig:paper2_element_embedding}) are also reproduced with the pre-trained model. In these cases, we also tried to extend the study to other materials, and all got reasonable results, so the claims in the paper are further confirmed. Interestingly, for the elemental embedding study, we also tried to use elemental symbols instead of names and got slightly different results (Figure \ref{fig:paper2_element_embedding_sub3}). It appears the cluster boundaries are not as clean as the one based on element name embedding (Figure \ref{fig:paper2_element_embedding_sub2}).

However, when we tried to train a new model with the package, we got an unexpected error that the model can not be trained. Further investigation showed that it was because one of the Mat2Vec's dependencies, gensim \cite{gensim},  has been updated and the latest version is not compatible with the infrastructure of Mat2Vec. Downgrading gensim to version 3.7.1 solved this issue. Moreover, the original dataset used to train the models is not provided, which prevented us from reproducing their Mat2Vec model from scratch, so we were unable to reproduce the last study of the work. Presumably, this is because of potential IP concerns or publisher agreements. The authors did, however, provide a comprehensive explanation of how they acquired and processed the training data. Also, the authors provided a training script with default training parameters included.

\begin{table}[!ht]

    \centering
    \caption{Reproduced results using the Word2Vec model for finding the most similar words given a reference word. The first column is the provided sample output using the word "thermoelectric". The second column is the reproduced results using the same word. The third column is the result with $band_gap$ as the reference phrase, which was not included in the original work}
    \resizebox{\textwidth}{!}{%
    \begin{tabular}{|c|cc|cc|cc|}
    \hline
        \textbf{Word} & \multicolumn{4}{c|}{thermoelectric} & \multicolumn{2}{c|}{band\_ gap} \\ \hline
        \textbf{} & \multicolumn{2}{c|}{Example provided} & \multicolumn{2}{c|}{Example replicated} & \multicolumn{2}{c|}{New examples}  \\ \hline
        \textbf{No.} & Output & Similarity & Output & Similarity &  Output & Similarity \\  
        \textbf{1} & thermoelectrics & 0.844 & thermoelectrics & 0.844 &  bandgap & 0.935 \\ 
        \textbf{2} & thermoelectric\_properties & 0.834 & thermoelectric\_properties & 0.834 & band\_-\_gap & 0.933 \\ 
        \textbf{3} & thermoelectric\_power\_generation & 0.793 & thermoelectric\_power\_generation & 0.793 &  band\_gaps & 0.861 \\ 
        \textbf{4} & thermoelectric\_figure\_of\_merit & 0.792 & thermoelectric\_figure\_of\_merit & 0.792 & direct\_band\_gap & 0.851\\ 
        \textbf{5} & seebeck\_coefficient & 0.775 & seebeck\_coefficient & 0.775 &  bandgaps & 0.819 \\ 
        \textbf{6} & thermoelectric\_generators & 0.7649 & thermoelectric\_generators & 0.764 &  optical\_bandgap & 0.814 \\ 
        \textbf{7} & figure\_of\_merit\_ZT & 0.759 & figure\_of\_merit\_ZT & 0.759 &  optical\_band\_gap & 0.813 \\ 
        \textbf{8} & thermoelectricity & 0.752 & thermoelectricity & 0.752 & band\_gap\_energies & 0.800 \\ 
        \textbf{9} & Bi2Te3 & 0.748 & Bi2Te3 & 0.748 &  direct\_bandgap & 0.788 \\ 
        \textbf{10} & thermoelectric\_modules & 0.743 & thermoelectric\_modules & 0.743 & eg & 0.787 \\ \hline
    \end{tabular}}
    \label{tab:paper2_simlarity}
\end{table}

Overall, this paper offers an impressive example of reproducibility, a practice from which the materials science community could learn. The authors have packaged their work in a tidy, comprehensive way that encourages further exploration and extension of their ideas. The code is not only well-documented, but also supported with tutorials that guide users through installation and model testing. We commend the authors for their attention to detail, as evidenced by the ease with which we were able to reproduce and expand upon the results and figures in the paper. However, like any exploratory work, there are areas that offer room for improvement. During our investigation, we encountered two minor issues. The first pertained to the versioning of Mat2Vec's dependencies, specifically gensim, and the second related to the absence of the training data. Reflecting upon this experience, we propose that: 1) It could be beneficial to stabilize the versions of dependencies rather than linking them to their most recent iterations, thereby circumventing potential compatibility issues. 2) Although the authors have justifiably omitted their training data, providing more detailed information about its acquisition and processing could bolster the interpretability and reliability of the results. A greater degree of transparency in this area, potentially extending to the sharing of processing scripts, would be advantageous.

Delving deeper into the second point, the concern here is not necessarily model accuracy but also the potential for bias in the trained model that might be unnoticed without access to the original data. The inability to peer-review the dataset and the training process means the community may inadvertently overlook biases. These could take the form of exclusion or selection bias, stemming from the exclusion of foreign language abstracts or certain types of articles. Similarly, a focus on abstracts related to inorganic materials and a high recall classifier may unintentionally introduce a confirmation bias. Furthermore, the tokenization process the authors adopted, which reportedly improved their results, could inadvertently result in algorithmic bias. Thus, the community could benefit from a broader discussion about the fairness, interpretability, and potential biases in machine learning models in materials science. 

\section{Discussion}
In this section, we expanded our reproducibility study by conducting a comparison of the output generated by the two NLP models mentioned in the respective papers. The aim of this comparative analysis is to gain a better understanding of the application domains of each model, despite the absence of fully open-sourced details for either model's hyperparameters and training datasets. Table \ref{tab:cross_paper_comparasion} listed the most similar compounds outputted using Model 2 (NLP model of paper 2) for "LiFePO4", an example chemical formula provided in Paper 1, Additionally, it lists the most similar words generated by Model 1 (NLP model from Paper 1) for the term "thermoelectric," which serves as an example word from Paper 2. We found that both Model 1 and Model 2 yielded reasonable output for the tested words from the other model, indicating some evident similarities. When considering the overlap between the top 10 sets of both models, there appears to be a significant degree of commonality, which is expected given that both models belong to the same family and are trained on corpora within the same domain. It is worth noting that the preprocessing stage, particularly the treatment of underscore-joined phrases, has a clear impact on "application" words like "thermoelectric." This provides strong evidence that text processing, even if it involves simple "cleaning up" techniques, can greatly influence end-to-end reproducibility. The choice of preprocessing method may depend on the specific use case.

Furthermore, the results obtained for chemical formulas are also interesting. Based on the similarity scores listed in Table \ref{tab:paper1_simlarity} and Table \ref{tab:cross_paper_comparasion}, it is likely that the results from Paper 2 are more precise at identifying the most similar chemical formula to LiFePO4, in terms of the properties of the materials. While it is acknowledged that some differences may arise due to variations in text sources (such as full text versus abstract) and the specific list of papers used, the normalization of formulas in Mat2Vec removes one source of variance. In Paper 1, potential noise might have been introduced by up-weighting formulas written in a more "canonical" form, which may result in missing some of the less common instances. It is difficult to determine if this truly impacts the disparities in the lists, but it is noteworthy that preprocessing choices, such as formula normalization, can influence the reproducibility and the compatibility of models across related works. We have also tried to reproduce other plots in the second paper with the model provided by the first paper, and the results are given in SI (Section \ref{sec:comparative_study}). Although they were not the intended applications of the model, the model performed reasonably well in giving similar embeddings to similar elements (Figure \ref{fig:paper1_cross_element_embedding}), as well as predicting similar relationships between similar concepts (Figure \ref{fig:paper1_cross_arithmetic}). 

\begin{table}[!ht]

    \centering
    \caption{Results from running the other paper's NLP model}
    \begin{tabular}{|c|cc|cc|}
    \hline
        \textbf{Word} & \multicolumn{2}{c|}{thermoelectric} & \multicolumn{2}{c|}{LiFePO4} \\ \hline
        \textbf{Model} & \multicolumn{2}{c|}{Model 1} & \multicolumn{2}{c|}{Model 2}  \\ \hline
        \textbf{No.} & Output & Similarity & Output & Similarity  \\  
        \textbf{1} & Thermoelectric & 0.727 & LFP & 0.874 \\ 
        \textbf{2} & pyroelectric & 0.605 & Li3O12P3V2 & 0.8594 \\ 
        \textbf{3} & thermoelectrical & 0.602 & LiMnO4P & 0.843 \\ 
        \textbf{4} & photovoltaic & 0.595 & Li4O12Ti5 & 0.828 \\ 
        \textbf{5} & optoelectronic & 0.587 & FeLi2O4Si & 0.828 \\ 
        \textbf{6} & multiferroic & 0.580 & CoLiO4P & 0.816 \\ 
        \textbf{7} & electrical & 0.567 & Li2MnO4Si & 0.804 \\ 
        \textbf{8} & TCO & 0.567 & LiMn2O4 & 0.793 \\ 
        \textbf{9} & NTC & 0.558 & LVP & 0.772 \\ 
        \textbf{10} & piezoelectric & 0.555 & FeO4P & 0.769 \\ \hline
    \end{tabular}
    \label{tab:cross_paper_comparasion}
\end{table}

Since the advent of the works reviewed, the field of Natural Language Processing has seen rapid advancements, with the integration of increasingly sophisticated NLP algorithms such as large language models (LLM) into materials science. This progression has, however, resulted in a diminishing ratio of studies that are readily reproducible. There are several challenges to reproducibility. One such challenge arises when these complex algorithms are trained on extensive datasets, which are often inaccessible due to publisher agreements. An example is the MatBERT, a transformer model specifically for the materials science domain, proposed by Trewartha et al. (2022), trained on a corpus of 2 million papers from materials science journals that are not publicly available \cite{trewartha2022quantifying}. Another challenge is the reliance on models or tools that may be proprietary or lack permissions for industrial use. As an illustration, Bran et al. (2023) developed an LLM tool to address reasoning-intensive chemical tasks. They acknowledged that the released package contains a limited set of tools, and the results it produces differ from those reported in the paper \cite{bran2023chemcrow}. Furthermore, training large models, especially LLMs, often requires substantial computational resources, which poses yet another obstacle to reproducibility. The emerging trend of lab automation has also added an additional layer of complexity to this issue, involving intricate lab equipment setups. For instance, Skreta et al. (2023) employed iterative prompting of LLM to operate lab robots \cite{skreta2023errors}. Reproducing such setups effectively demands that researchers be well-versed in NLP and possess comparable resources. These limitations in resource access, combined with algorithmic complexity and the absence of open data, raise a crucial question: What aspects of the research can be reproduced under these constraints? Understandably, the more complex an algorithm, the less likely it is to be replicated. Nonetheless, the cornerstone of scientific integrity is the ability to reproduce and verify results. To promote positive progress in the material science NLP domain, it is our recommendation that future publications make their methods section as transparent as possible. No details should be left out, especially when a complex algorithm is used. Emphasizing reproducibility in upcoming studies will help to build a more robust and reliable scientific community.

\section{Conclusion}
Both studies provided well-written manuscripts describing their workflows in adequate detail and clean and well-documented codebases. Their repositories also have clear guidance from installing the packages, loading pre-trained models, and evaluating the outcome of the models to reproducing the results in the papers. With the information provided, we could replicate and reproduce most of the results shown in both articles. Further, we were also able to extend the analysis to other materials and further confirmed the validity of the models and claims in the papers. Therefore, future research publications in materials science and related fields could take inspiration from their practices for improved reproducibility.

However, despite the earnest efforts of the authors of both papers, we also encountered a few general challenges in our study. First, the training data from the published scientific articles were not open-sourced due to copyright restrictions. Second, the model training process remains a black box to the readers as the models were not released. Also, some details of the model's architecture and training procedure are missing. Both factors prevented future researchers from reproducing the works from scratch or fine-tuning the model for other application domains. However, these two observations have been common in NLP research beyond materials science. Finally, we have observed some backward compatibility issues due to updates in the dependencies, which could be circumvented by better specifying the versions of all underlying dependencies.

Natural Language Processing has seen rapid advancements, integrating increasingly complex NLP algorithms into materials science. This progression, however, has led to a decrease in the ratio of studies that can be easily reproduced. Challenges include large datasets often inaccessible due to publisher agreements, models or tools that may be proprietary or lack permissions for industrial use, and substantial computational resources required for training large models. Reproducing complex lab automation setups, which include intricate lab equipment, also poses an additional challenge. In light of these considerations, we recommend that future publications prioritize transparency, especially when complex algorithms are used. Every detail in the methods section should be disclosed and explained as clearly as possible. This emphasis on reproducibility will be instrumental in fostering a more robust and reliable scientific community.

Finally, we extend our acknowledgment to the authors of the two papers analyzed in this study for their considerable contributions to the field of materials science. Despite the challenges, their work stands as a testament to the power of meticulous research and transparent reporting. We encourage the research community to continue striving for transparency and reproducibility, strengthening the collective scientific enterprise.

\section{Author Contributions}
XL and SS conducted the experiments. All authors contributed to writing and editing the paper. EK was not involved in the direct execution of any reproducibility experiments, as he was an author on the first paper discussed in this work.

\bibliographystyle{unsrt}  
\bibliography{references}  

\newpage 

\setcounter{section}{0} 
\setcounter{equation}{0} 
\setcounter{figure}{0} 
\setcounter{table}{0} 

\renewcommand{\thefigure}{S\arabic{figure}} 
\renewcommand{\thetable}{S\arabic{table}} 
\renewcommand{\theequation}{S\arabic{equation}} 
\renewcommand{\thesection}{S\arabic{section}} 

\section{Supplementary Information}
\subsection{Comparative studies}
\label{sec:comparative_study}
Attempts to reproduce studies mentioned in paper 2 with the Word2Vec model provided in paper 1. We note that these applications were different from the intended applications for the model, and none of these studies were reported in the original publication.

\begin{figure}[ht]
    \centering

    \begin{subfigure}{0.45\textwidth}
        \centering
        \includegraphics[width=\textwidth]{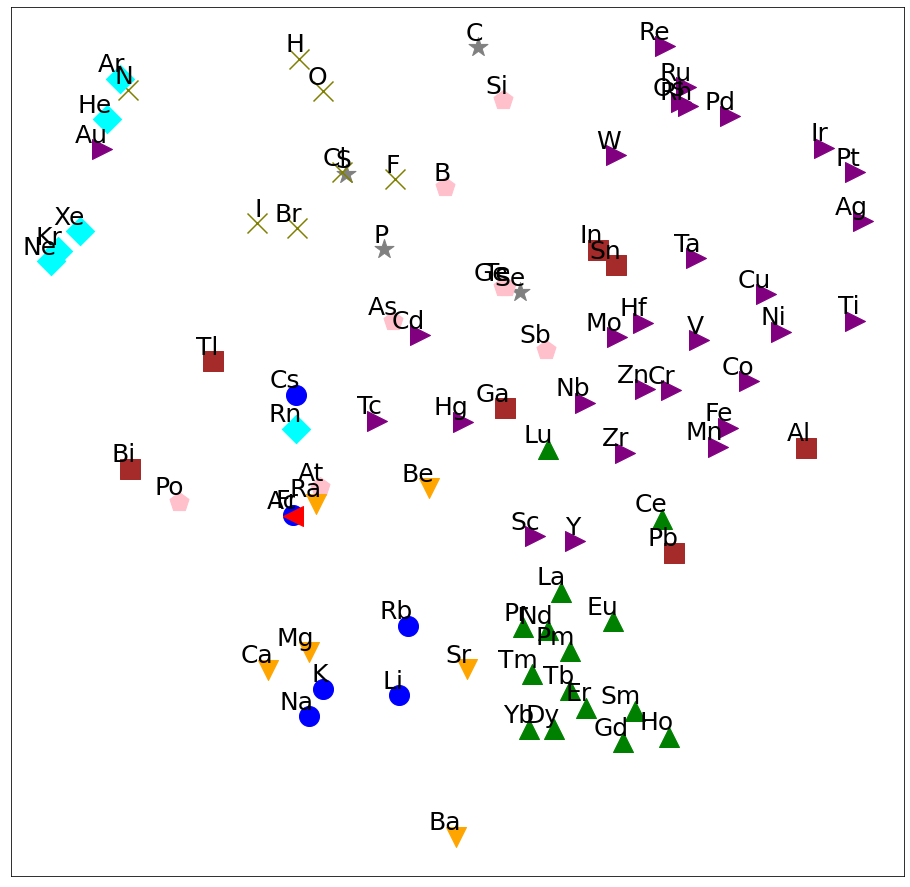}
        \label{fig:paper1_cross_element_embedding_sub1}
    \end{subfigure}
    \begin{subfigure}{0.45\textwidth}
        \centering
        \includegraphics[width=\textwidth]{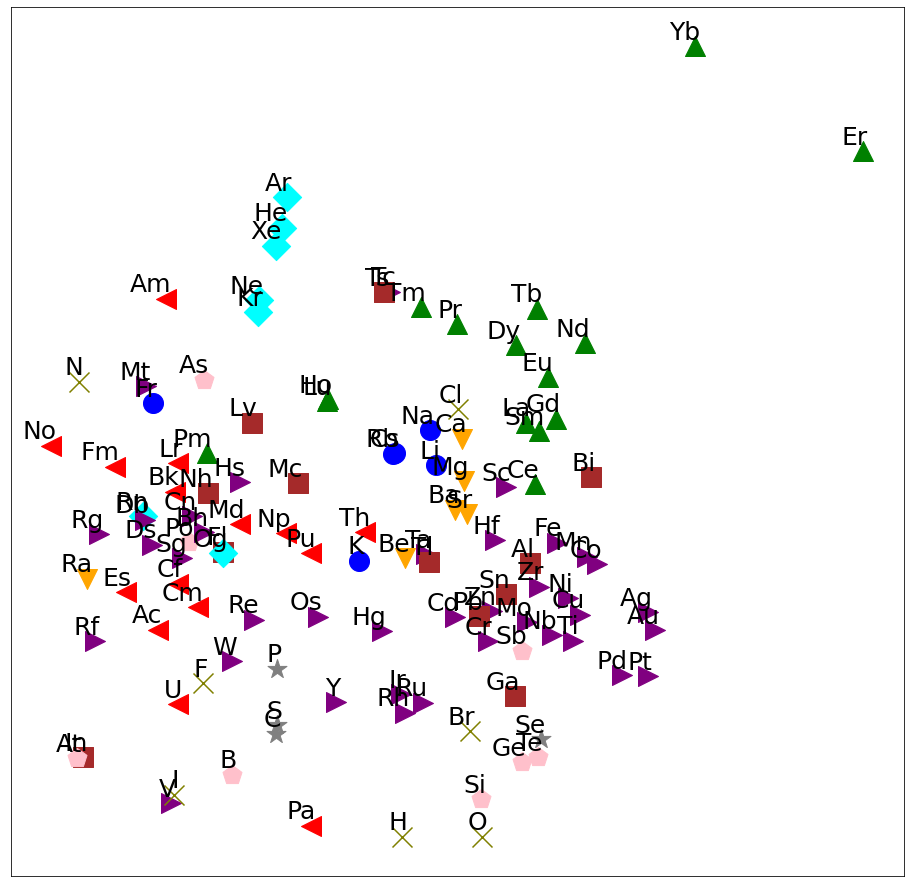}
        \label{fig:paper1_cross_element_embedding_sub2}
    \end{subfigure}
    \caption{Study of elemental embeddings using the Word2Vec model provided by paper 1. Same plot as Figure \ref{fig:paper2_element_embedding}. a) Two-dimensional t-distributed stochastic neighbor embedding (t-SNE) projection of the word embeddings of 100 chemical element names (for example, ‘hydrogen’) labeled with the corresponding element symbols. b) Same plot but based on the embeddings of element symbols}
    \label{fig:paper1_cross_element_embedding}
\end{figure}

\begin{figure}[ht]
    \centering

    \begin{subfigure}{0.6\textwidth}
        \centering
        \includegraphics[width=\textwidth]{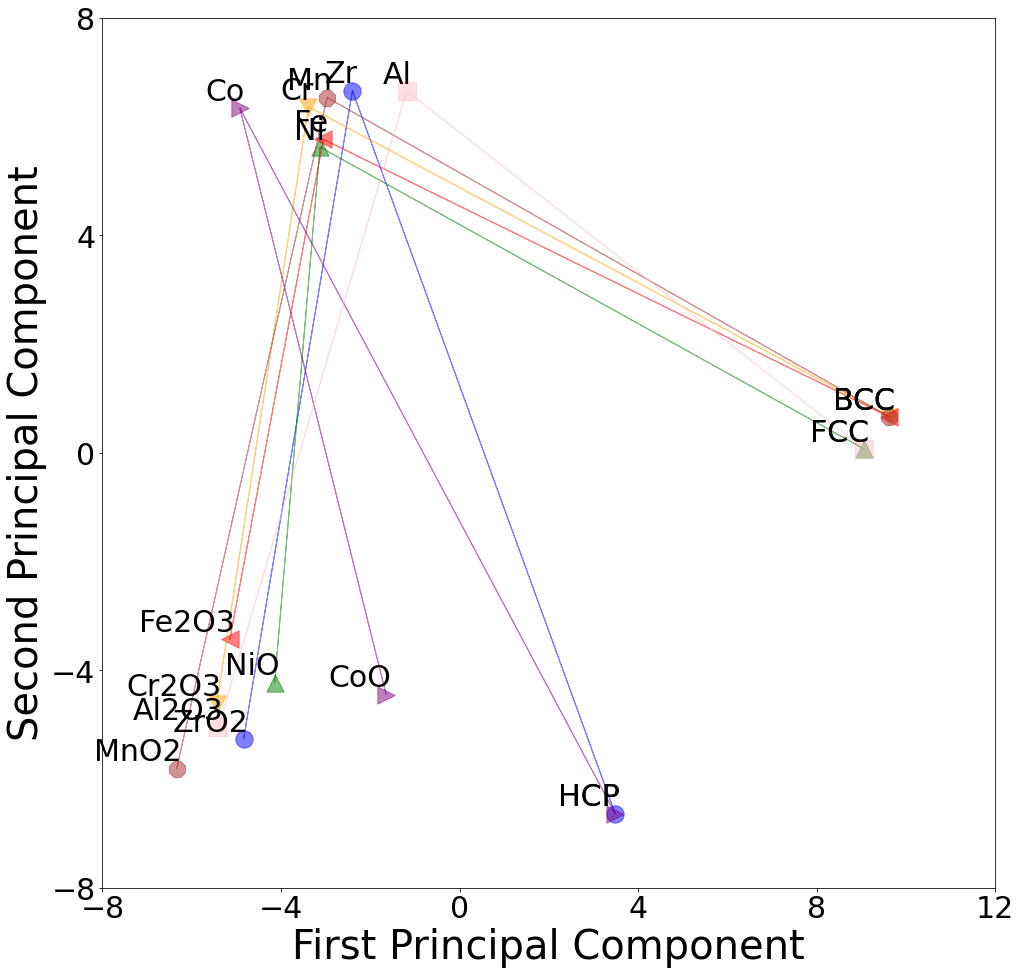}
        \label{fig:paper1_cross_arithmetic_sub1}
    \end{subfigure}
    \caption{study of word relationships based on the predicted word embeddings using Word2Vec model provided by paper 1. Same plot as Figure \ref{fig:paper2_arithmetic}.  Word embeddings for Zr, Cr and Ni, their principal oxides, and crystal symmetries (at standard conditions) projected onto two dimensions using principal component analysis and represented as points in space. The relative positioning of the words encodes materials science relationships, such that there exist consistent vector operations between words that represent concepts such as ‘oxide of ’ and ‘structure of’.}
    \label{fig:paper1_cross_arithmetic}
\end{figure}

\end{document}